\documentclass[twocolumn,prl,showpacs,preprintnumbers,amsmath,amssymb]{revtex4-1}
\usepackage{graphicx}
\usepackage{dcolumn}
\usepackage{bm}
\usepackage{longtable}
\usepackage{epsfig}

\begin{document}

\title{Oxidation mechanism in metal nanoclusters: Zn nanoclusters to ZnO hollow nanoclusters}

\author{A K Mahapatra}
\email{amulya@iopb.res.in (A K Mahapatra) }
\author{U M Bhatta}
\author{T Som}
\email{tsom@iopb.res.in (T Som)}
\affiliation{Institute of Physics, Sachivalaya Marg, Bhubaneswar 751005, India}
 
\begin{abstract}

Zn nanoclusters (NCs) are deposited by Low-energy cluster beam deposition technique. The mechanism of oxidation is studied by analysing their compositional and morphological evolution over a long span of time (three years) due to exposure to ambient atmosphere. It is concluded that the mechanism proceeds in two steps. In the first step, the shell of ZnO forms over Zn NCs rapidly up to certain limiting thickness: with in few days --- depending upon the size --- Zn NCs are converted to Zn--ZnO (core--shell), Zn--void--ZnO, or hollow ZnO type NCs. Bigger than $\sim$15 nm 
become Zn--ZnO (core--shell) type: among them, NCs above $\sim$25 nm could able to retain their initial geometrical shapes (namely triangular, hexagonal, rectangular and rhombohedral), but $\sim$25 to 15 nm size NCs become irregular or distorted geometrical shapes. NCs  between $\sim$15 to 5 nm become Zn--void--ZnO type, and smaller than $\sim$5 nm become ZnO hollow sphere type i.e. ZnO hollow NCs. In the second step, all Zn--void--ZnO and Zn--ZnO (core--shell) structures are converted to hollow ZnO NCs in a slow and gradual process, and the mechanism of conversion proceeds through expansion in size by incorporating ZnO monomers inside the shell. The observed oxidation behaviour of NCs is compared with theory of Cabrera -- Mott on low-temperature oxidation of metal. 
\end{abstract}

\maketitle
\section{Introduction}

Future generation devices will use a lot of nanometer-size metals in them. Nanomaterials have a large surface-to-volume ratio, and enthalpy of oxide 
formation for many metals is low \cite{ch5campbell}. Hence, nanometals easily get oxidized, even in a vacuum environment. Functioning and life time of
 devices could be significantly altered if its nanometal components get oxidized. Therefore, it is imperative to understand the mechanism of oxidation in metals of nanometer size. It can also be noted that metal--oxide (metal--semiconductor) interfaces have been playing a key role in many micro-electronics and opto-electronic applications \cite{ch5sze}. Hence, fabrication of metal--oxide interface in nanometer scale by a simple process of oxidation in a controlled manner could be very helpful in technological applications as well as for fundamental studies.

Oxidation process on bulk metal surfaces has been studied extensively for a century. The process of oxidation at low 
temperature is generally believed to follow in accordance with the mechanism proposed by Cabrera and Mott (CM). In CM 
theory \cite{ch5cabrera} it is assumed that a layer of atomic oxygen adsorbs on a metal surface when it gets exposed to 
atmosphere. Electrons from metal can tunnel through the initial oxide layers and ionize the adsorbed oxygen atoms. It results in setting up a potential difference across the oxide layer. The potential difference lowers the activation energy required for metal ions to diffuse and helps in increasing the oxide thickness at oxide-gas interface. The growth rate of oxide is very high up to a certain limiting thickness after which it becomes so low that an increase in oxide thickness practically stops. Hence, at constant temperature, thin uniform layer of oxide forms over a metal surface and helps in protecting the metal from further degradation.

Study on oxidation behaviour of metals at nanometer scale is comparatively new \cite{ch5niklasson,ch5ogata,ch5rao,ch5tamura}. Morphological and compositional evolution of few metals, including Zn, of nanometer size by annealing in ambient atmosphere for short period of time (a few minutes), has been studied in recent years \cite{ch5nakamura1,ch5nakamura2} -- though not in the atomic resolution level. However, the evolution of any metal NCs, simply due to exposure to ambient atmosphere, without annealing, is not yet reported. In this paper we are reporting it for Zn NCs with an atomic resolution by exposing the samples to ambient atmosphere over a long span of time (three years): it gives invaluable insight into the mechanism of oxidation in metal NCs. In addition, two major concerns related to oxidation process are addressed: first, effect of shape and size on the process; and second, adequateness of CM theory in explaining the process as size scales down to nanometer range. For the purpose, Zn NCs are produced by low-energy cluster beam deposition (LECBD) technique. Needless to mention, ZnO is one of the most studied material for its possible potential application as solar cell \cite{ch5kevin}, transparent conducting film \cite{ch5islam}, diluted magnetic semiconductor \cite{ch5zhao,ch5ueda}, optoelectronic \cite{ch5huang}, piezoelectric material \cite{ch5wang} and so on. 

The LECBD technique has many advantages over other techniques for this specific study. The unique advantage of LECBD technique is NCs form through homogeneous nucleation process: the effect of substrate on shape is insignificant as NCs first form in a `flight-condition' and then deposit on substrate in a `soft-landing' process. Hence,  it is possible to deposit well separated NCs with varied but well-defined geometrical shapes in a single run. It helps in studying  oxidation behaviour of nanometals for varied size, and most importantly for varied shape, by analysing each NC independently by high resolution transmission electron microscopy (HRTEM).

It is concluded that oxidation mechanism proceeds in two steps: In first step, the shell of ZnO forms over Zn NCs rapidly up to certain limiting thickness; and in the next step, the shell swells gradually in a very slow rate by incorporating ZnO monomers inside it without increasing the thickness, which leads to instability and breakage. In the initial step of oxidation, the final morphology and composition depend on the initial size of Zn NCs and the big size NCs could retain their original shape. But, in the next step of oxidation Zn NCs of all sizes and shapes are converted to hollow ZnO NCs. Generally, voids are not spherical in shape and related to initial shape of Zn NCs.     
 
\section{Experimental}

Zn NCs are deposited by LECBD technique. Deposition of Zn NCs by the LECBD technique can be considered as a three step process: (i) formation of Zn atomic vapours, (ii) homogeneous nucleation and growth of NCs in an argon  gas atmosphere and (iii) extraction (from growth region) and deposition of NCs on the substrate. The cluster generating source used for this  study is of the Sattler type \cite{ch5sattler}. Zn vapours are formed by heating Zn powder using resistive coil in a molybdenum crucible at 740 $K$. Zn vapours come out of the crucible through a hole that is present in the side wall. A carrier gas (namely argon) continuously flows into the condensation chamber and gets out through a tubular outlet to the vacuum chamber. This tubular outlet is kept at a low temperature by circulating liquid nitrogen ($LN_2$) around it. Zn monomers transfer their energy to $LN_2$-cooled inert gas atoms and achieve supersaturation \cite{ch5min}. This results in homogeneous nucleation and growth of Zn NCs. The cluster generating source and substrates is kept in a vacuum chamber where the pressure is maintained at $\sim10^{-6}$ mbar. Passage of  gas leads to a rise in the chamber pressure to $\sim$ $10^{-4}$ mbar. Thermophoretic force \cite{ch5han} originating from temperature gradient between the hot crucible and the $LN_2$-cooled tubular outlet, and force due to  pressure gradient help NCs to come out of the gas condensation chamber through the tubular outlet. NCs coming out of the tubular outlet undergo an adiabatic expansion which leads to further reduction in their temperature. Since NCs go through successive stages of cooling, they have very low  kinetic energy and  get deposited on the substrate without fragmentation. Transmission electron microscope (TEM) Cu-grids attached on silicon wafers are
used as the substrate. Details of the LECBD set-up have been reported elsewhere \cite{ch5ravi}. After deposition,  flow of argon gas and $LN_2$ is stopped. Pressure of chamber again comes back to $\sim10^{-6}$ mbar after stopping flow of argon gas and this condition is maintained for around 10 h. During this time, temperature of the crucible slowly cools down to room temperature. Then the chamber pressure is maintained at $\sim10^{-3}$ mbar for another one day before breaking the vacuum of the chamber. The TEM grids are preserved inside an air-tight container after removing from deposition chamber. TEM is performed using JEOL-2010 operated at 
200 KeV electron beam energy. The exposure to electron beam does affect oxidation behaviour of metals \cite{ch5zhukov}; therefore, it is the aim to maintain the duration and frequency of TEM measurements minimum: the experiment is carried out after ten days (sample 1), three months, and then after three years (sample 2). 

\section{Results and discussion}
                       
\begin{figure*}
\begin{center}
\includegraphics*[width=7.0cm]{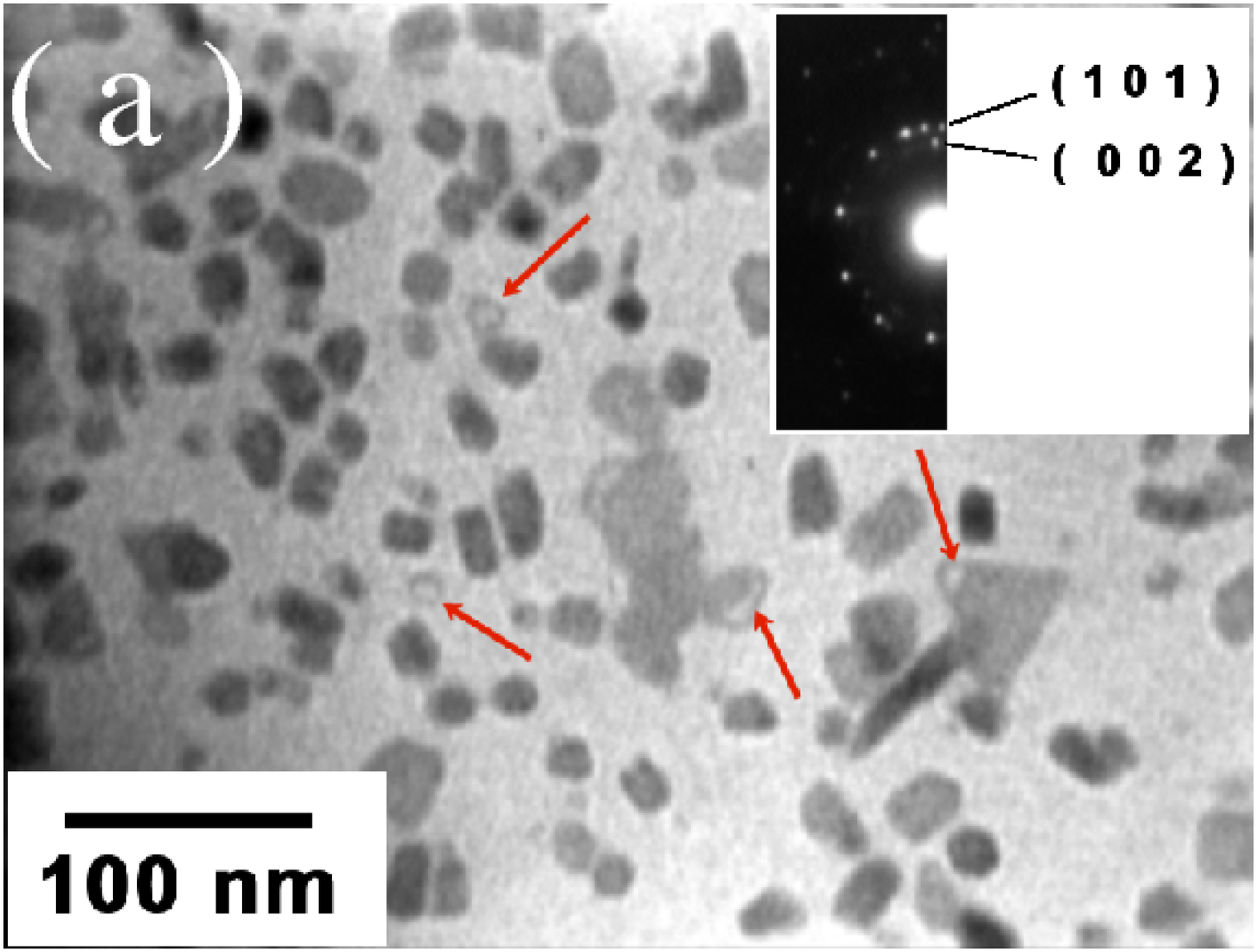}
\includegraphics*[width=7.0cm]{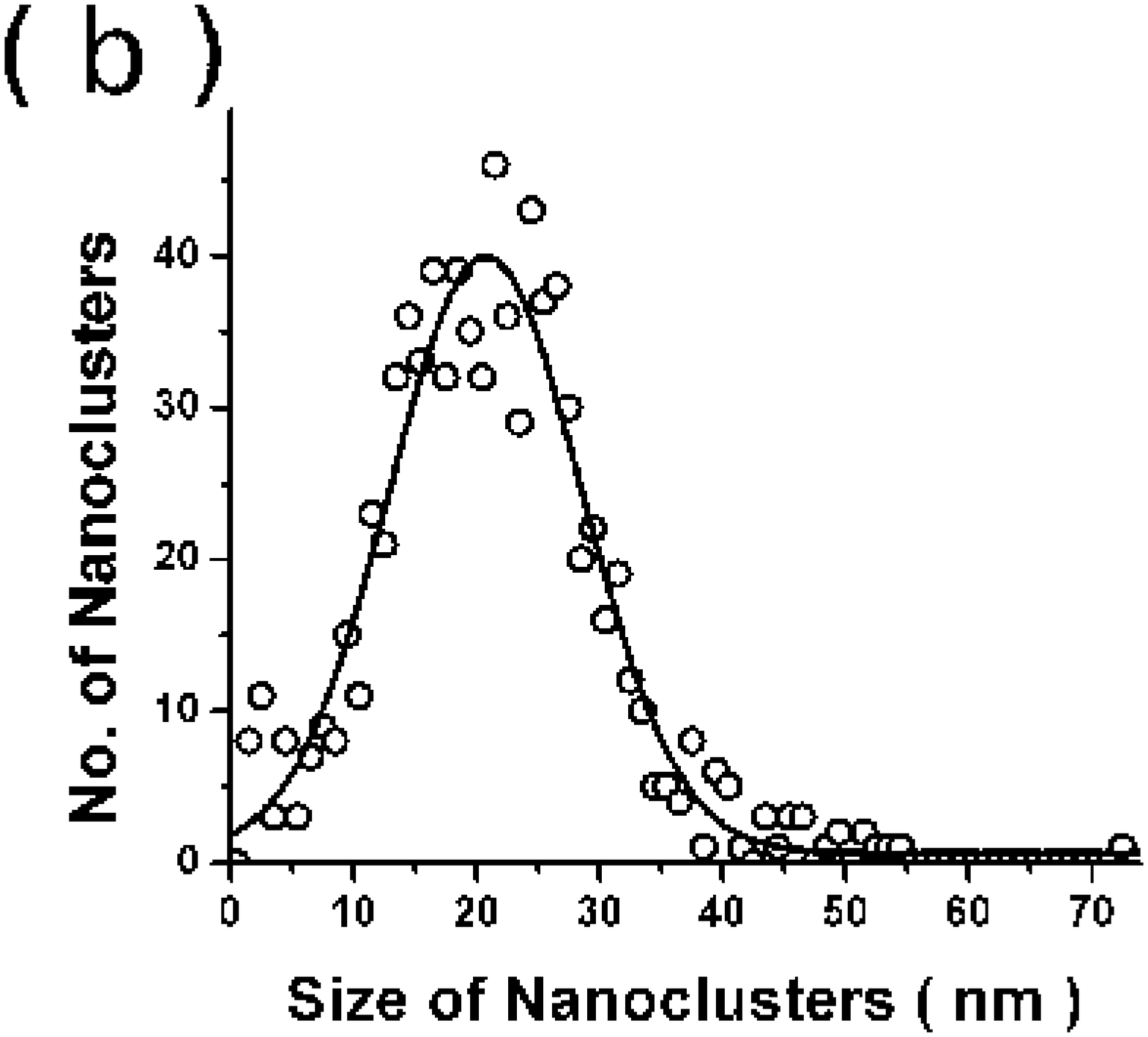}
\includegraphics*[width=7.0cm]{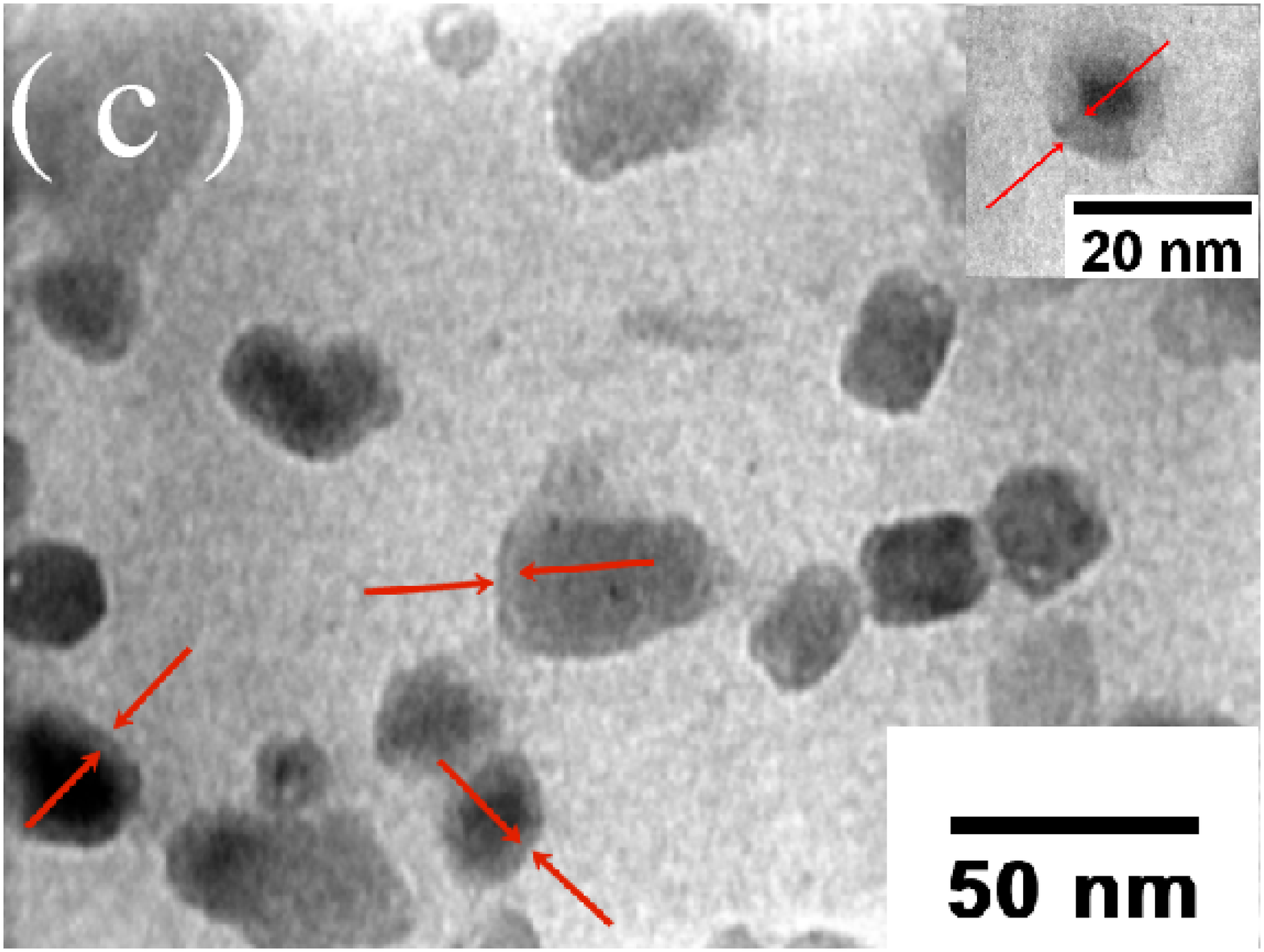}
\includegraphics*[width=7.0cm]{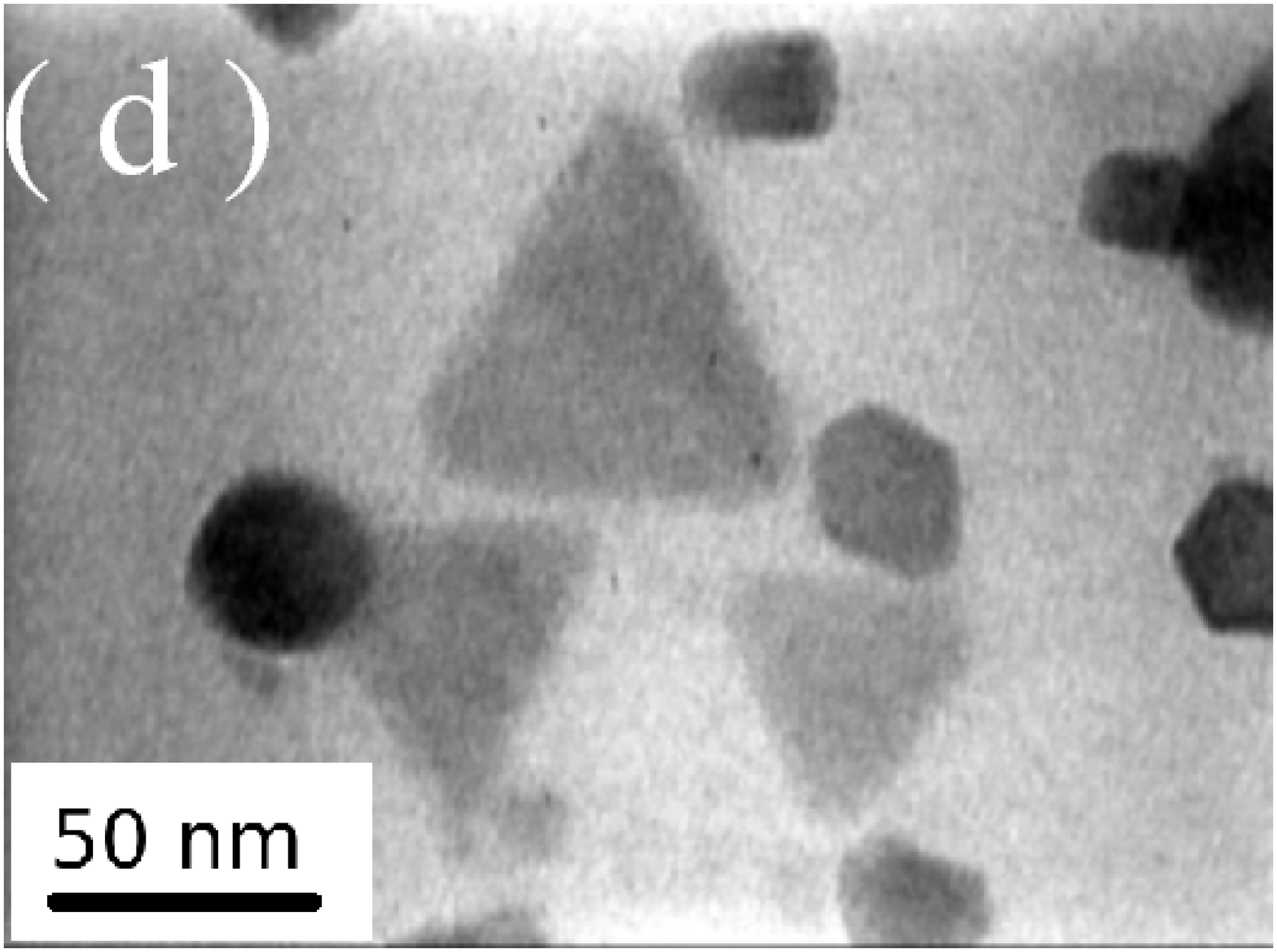}
\includegraphics*[width=7.0cm]{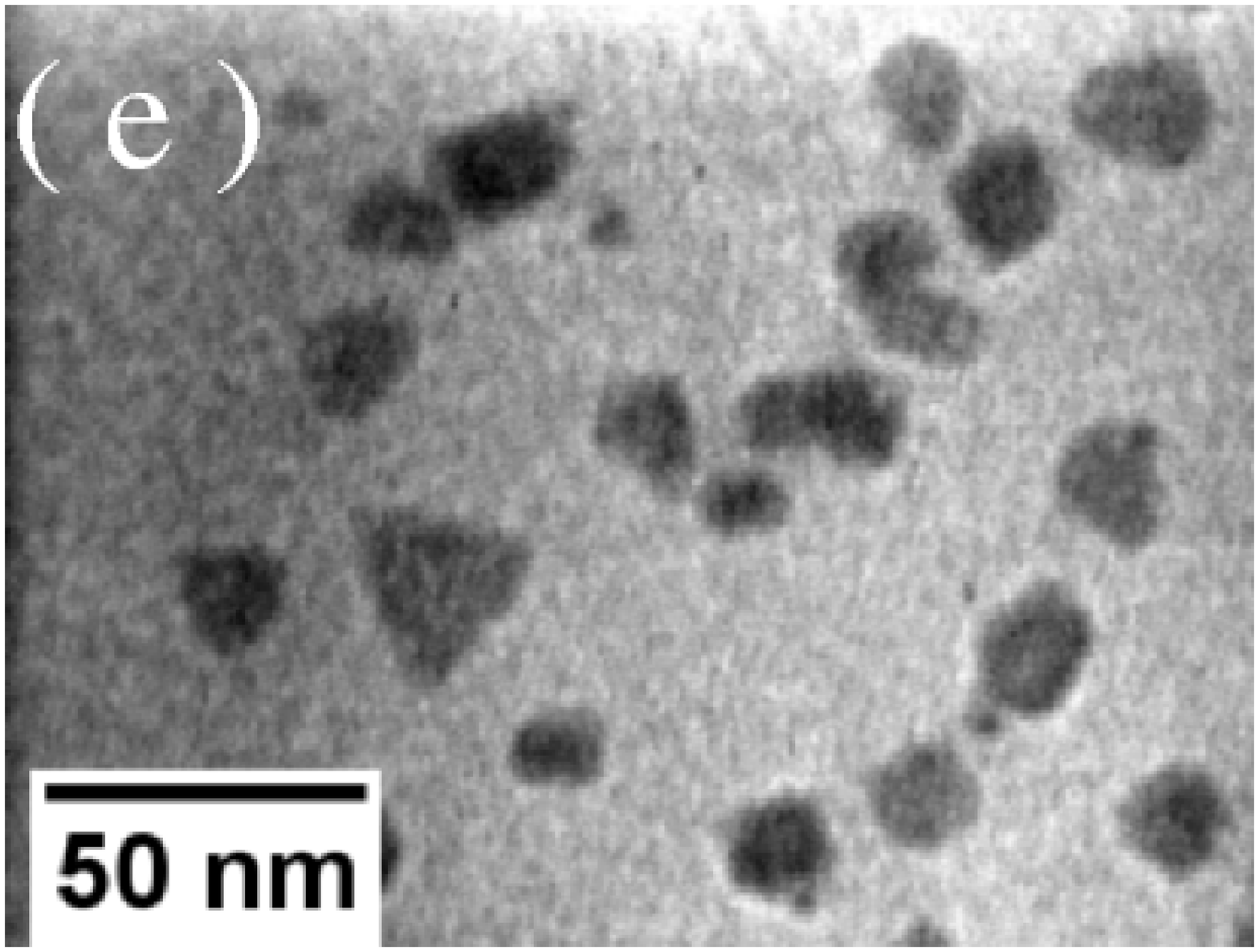}
\includegraphics*[width=7.0cm]{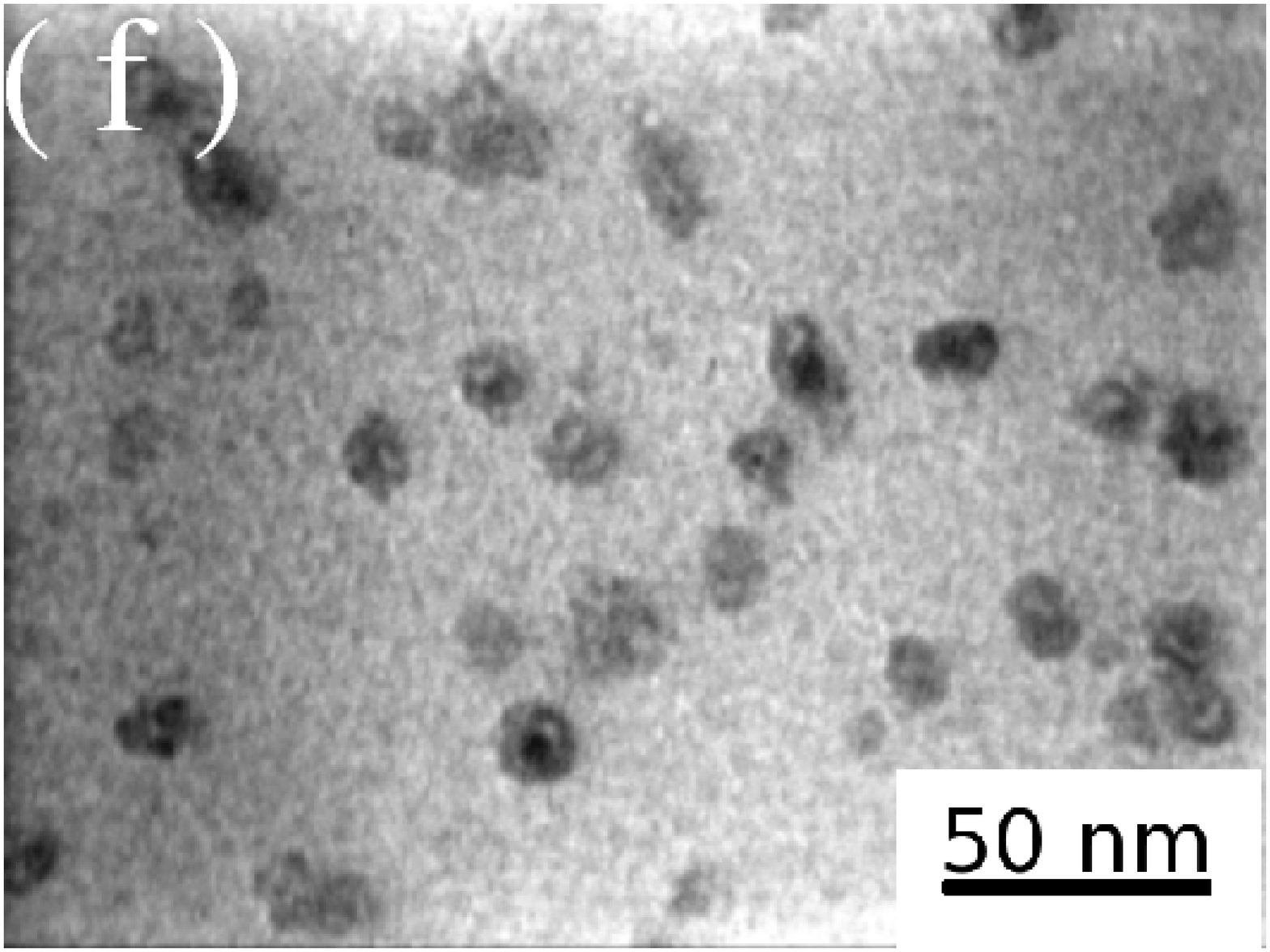}
\end{center}
\vspace{-0.5 cm}
\caption{{ Typical TEM micrographs of Zn/ZnO NCs (sample 1): (a) the arrow marks highlight the presence of voids in the NCs (SAED pattern is shown in the inset), (b) size distribution of NCs, 
(c) NCs of different size group, where arrow marks show the ZnO shell with a lighter contrast. A single Zn--ZnO core--shell NC, at a higher magnification, is shown in the inset, (d) geometrical shaped NCs of average size larger than $\sim$25 nm, (e) irregular or distorted geometrical shaped NCs of size $\sim$25 to 15 nm, and
 (f) Zn--void--ZnO type NCs of size $\sim$15 to 5 nm.}}
\label{ch5fig1}
\end{figure*}

Typical TEM micrographs of sample 1 are shown in figure \ref{ch5fig1}. The projected area diameter of several NCs is measured and the number of NCs counted per each size range is plotted in figure \ref{ch5fig1}(b). A Gaussian distribution with an average size of 20.7 nm and standard deviation of 7.9 nm best fits to the data points. Selected-area electron diffraction (SAED) pattern of NCs is shown in the inset of figure \ref{ch5fig1}(a). The $d$-spacings measured from (SAED)  is in good agreement with those of the (0 0 2) and (1 0 1) planes of Zn, corresponding to the hexagonal crystal structure (space group = P$6_{3}/mmc$) with lattice parameters $a$ = 0.2665 nm and $c$ = 0.4947 nm.  

\begin{figure*}
\begin{center}
\includegraphics*[width=15.0cm]{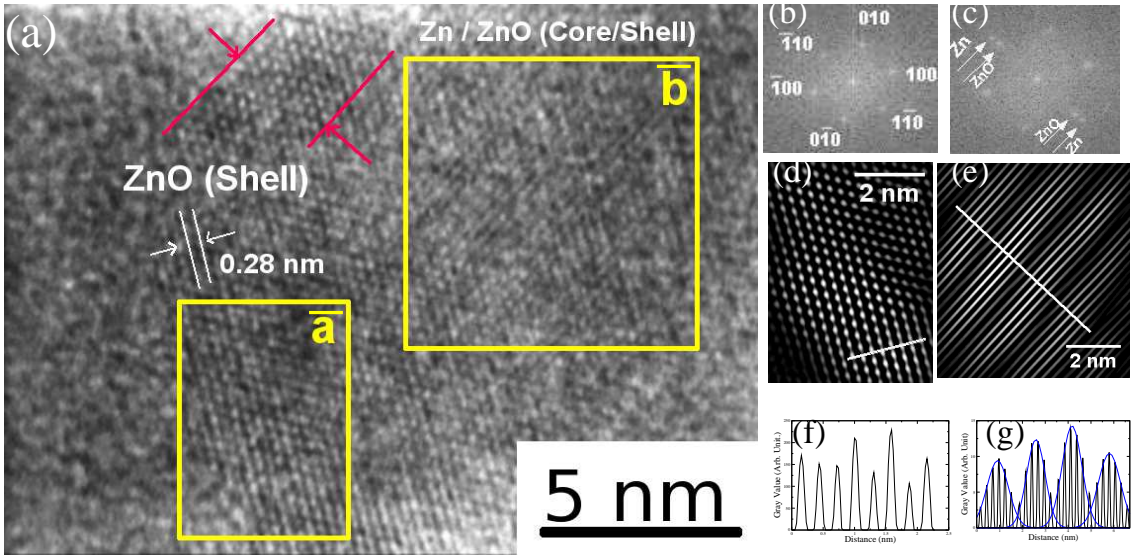}
\end{center}
\caption{ {(a) Typical HRTEM image of a hexagonal shaped Zn--ZnO core--shell NC, (b) FFT pattern of the shell region, that is outlined by a rectangle (\={a}) in (a), (c) FFT pattern of the core region, that is outlined by rectangle (\= {b}) in (a), (d) reconstructed HRTEM after filtering the FFT spot patterns of (b), (e) reconstructed HRTEM image after filtering the FFT spot patterns, marked with arrows, of (c), (f) grey value along the line drawn in (d), and (g) grey value along the line drawn in (e).}}
\label{ch5fig2}
\end{figure*}

A contrast between ZnO shell (lighter contrast) and Zn core (darker contrast) of the NCs is observed in  
TEM micrographs when viewed with higher magnification (marked by arrows in figure \ref{ch5fig1}(c)). Contrast between the core and the shell region becomes more prominent with further magnification. Zn--ZnO core-shell structure of a single  NC is shown in the inset of figure \ref{ch5fig1}(c). HRTEM (figure \ref{ch5fig2}(a)) validates the formation of Zn--ZnO 
(core-shell) NC. ZnO (shell) region is marked in the figure and the thickness measured to be around 3.0 nm. The fast Fourier transform (FFT) pattern of shell region (rectangular outline (\= {a}))  is shown in figure \ref{ch5fig2}(b). The spots are indexed corresponding to hexagonal crystal structure (space group = P$6_{3}mc$) having lattice parameters $a$ = 0.324 nm and $c$ = 0.5206 nm. The spots correspond to (1 0 0) family of lattice planes along the [0 0 1] zone axis . By filtering the FFT spot patterns, the original HRTEM of region (a) in figure \ref{ch5fig2}(a) is reconstructed and is shown in figure \ref{ch5fig2}(d). The $d$-spacing of 0.28 nm is obtained from the line profile study (figure \ref{ch5fig2}(f)), which corresponds to the (1 0 0) planes of ZnO. The HRTEM image shown here is a portion of a hexagonal shaped Zn-ZnO (core-shell) NC. The Zn NC is faceted with (1 0 0) planes over which ZnO (1 0 0) plane grows epitaxially. This results in observation of Moir\'{e} pattern in the core region as d-value of Zn and ZnO are different (rectangular outline (\= {b}) in 
figure \ref{ch5fig2}(a)). FFT of the region clearly shows that the spots correspond to the (1 0 0) planes of ZnO and Zn  along the [0 0 1] zone axis (marked by arrows in figure \ref{ch5fig2}(c)) and also suggests $(100)_{Zn}|| (100)_{ZnO}$ epitaxial relationship \cite{ch5ding}. To highlight the Moir\'{e} fringes, the spots marked with arrows in FFT, are filtered and the HRTEM image is reconstructed (figure \ref{ch5fig2}(e)). The line profile is plotted in figure \ref{ch5fig2}(g). It may be noticed that in each Moir\'{e} repeat six planes of ZnO are present and the spacing between Moir\'{e} repeat is 1.6 nm. If $d_{shell}$ and $d_{core}$ are the inter-planar spacings of the shell and the core regions then  spacing between the Moir\'{e} fringes will be $ D={d_{shell}d_{core}}/{(d_{shell}-d_{core})}$ and the number of planes 
in a Moir\'{e} repeat is given by $n={d_{shell}}/{(d_{shell}-d_{core})}$. By putting $d$-spacing of the (1 0 0) planes of both Zn and ZnO in the above equation $D$ = 1.3 nm and $n$ = 5.6 are obtained, which is consistent with the observed value. 

The enthalpy of oxide formation is quite low for zinc \cite{ch5campbell}. Hence, zinc oxidizes easily when it comes in contact
with oxygen. As per the CM theory, the oxide thickness increases according to the following growth rate \cite{ch5cabrera}:

\begin{equation}
\frac{dX}{dt}= N'\Omega \nu exp\biggl(\frac{qa'V}{kTX}-\frac{W}{kT} \biggr)
\label{ch5cabrera}
\end{equation}

where, $a'$ is the interatomic distance in  oxide, $q$ is the charge on ion, $V$ is the potential difference developed across the layer, $N$ is the number of  atoms per unit area in the metal surface, $\Omega$ is the volume of oxide per metal ion, 
$\nu$ is the vibrational frequency of atoms at the interface, $W$ is the total potential energy for incorporation  and subsequent migration of defect inside the oxide, $k$ is the Boltzmann constant, and $T$ is the absolute temperature. To compare  experimental results, an attempt has been made to estimate  limiting thickness of the oxide shell by taking standard values of  parameters available in the literature \cite{ch5CRC}: $a'$ = 0.25 nm, 
$q$ = +2e, $N' = 1.6\times 10^{15}$ atoms cm$^{-2}$, $\Omega = 2.3\times10^{-23}$ cm$^3$ atom$^{-1}$,
 $\nu = 4.3\times10^{12}$ s$^{-1}$. 
At room temperature, the value of $kT$ can be taken as 0.025 eV.  The value of $V$ can be approximated to 
$ {-\Delta G(ZnO)}/{2e} $, where $\Delta G$ is the free energy for formation of the oxide per oxygen atom \cite{ch5atkinson}. Taking $\Delta G$(ZnO) = --318.2 kJ mol$^{-1}$, potential difference developed across the thin film (V) will be 1.65 Volt. The value of $W$ can be approximated to 
the activation energy of the metal tracer diffusion coefficient in the oxide and is taken as 
1.8 eV \cite{ch5zwuensch}. Putting these values in equation (\ref{ch5cabrera}), it can be seen that at oxide thickness of 
1.1 nm, which forms within few hours, the growth rate becomes
0.1 nm per day and finally reduces to 0.1 nm  per year corresponding to an oxide thickness of 1.3 nm only. It means, the thickness of ZnO shell is expected to be $\leqslant$ 1.3 nm. However, oxide shell thickness even up to 5 nm is observed in some portions of NCs, which is much larger than the value estimated from equation (\ref{ch5cabrera}).  Nakamura {\it et al.} also observed a ZnO shell thickness of 4 nm  
over Zn NCs at room temperature \cite{ch5nakamura2}. 

\begin{figure*}
\begin{center}
\includegraphics*[width=7.0cm]{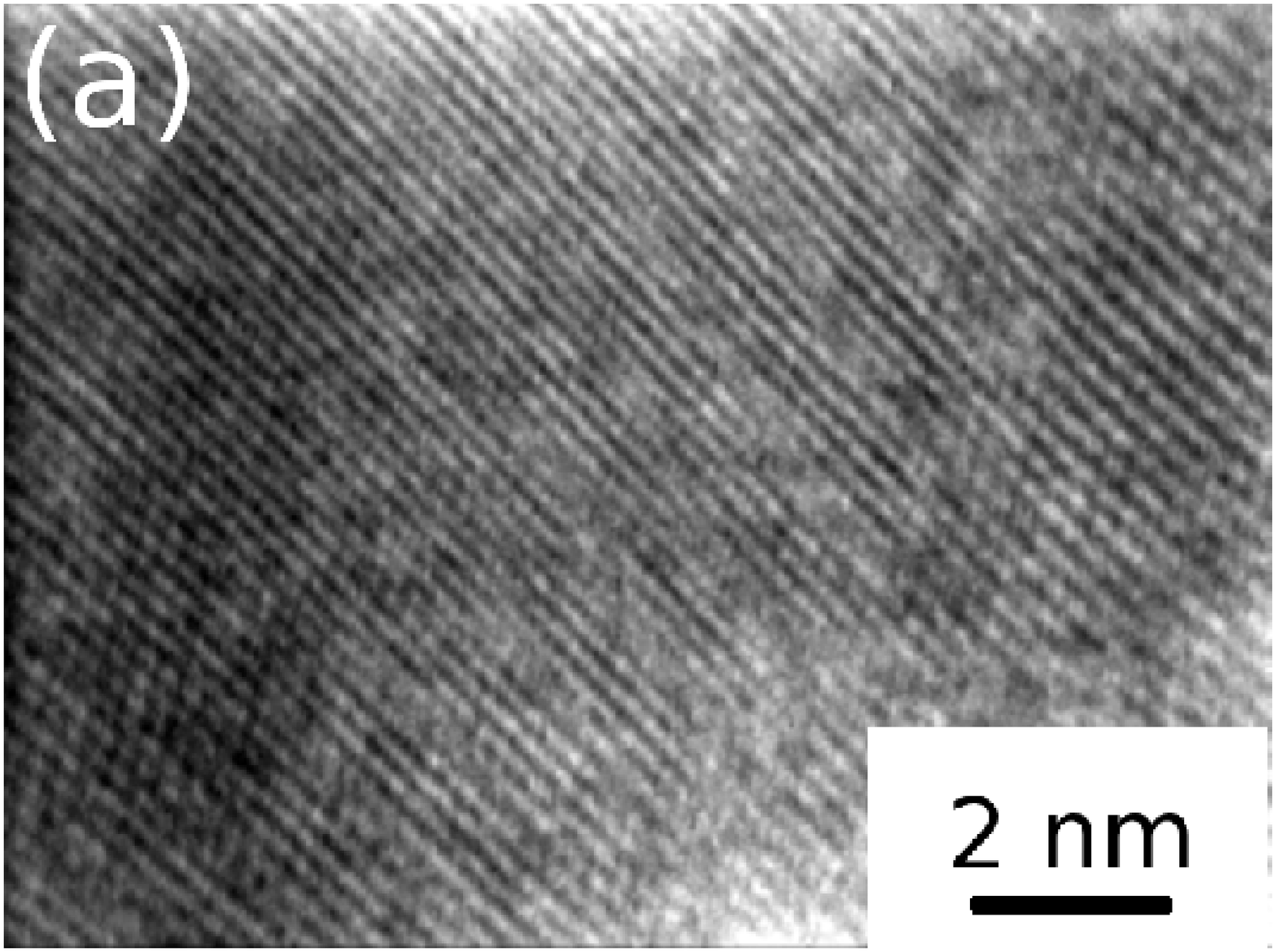}%
\includegraphics*[width=7.0cm]{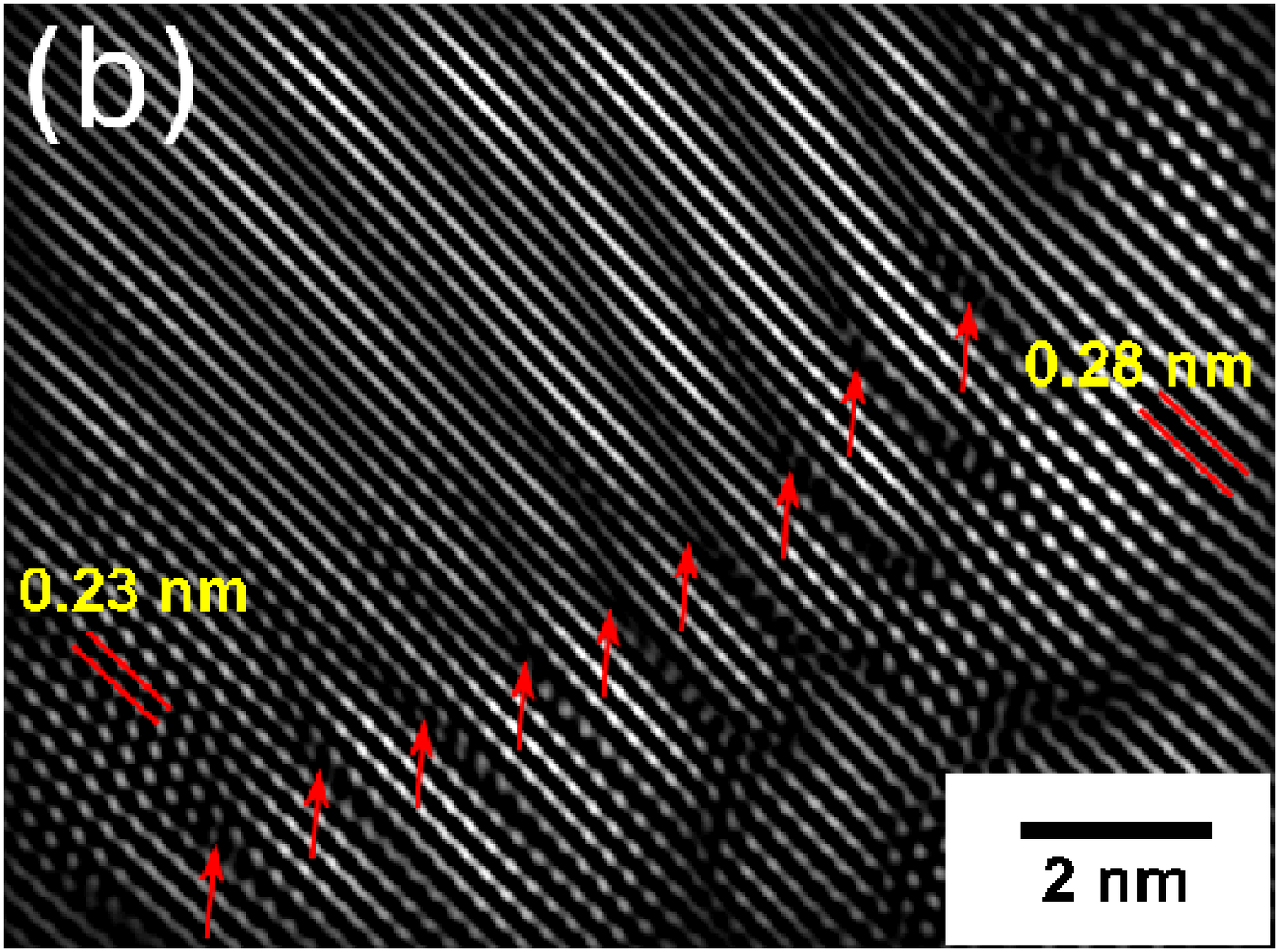}
\end{center}
\caption{{(a) HRTEM of interface showing the presence of dislocations, (b) reconstructed image after filtering FFT pattern of HRTEM of (a).}}
\label{ch5dislocation}
\end{figure*}

The observed oxide shell thickness is much larger than the estimated value possibly due to following reasons:
(i) when size of a material reduces to nanometer range, electronic property of the material changes. Hence, the value of $V$, basically the difference in metal Fermi level and  $O^-$ level, would be  different than the one used in the equation. In principle, the value of $V$ will be different from one NC to the other, depending on its shape and size. 
(ii) In CM theory, it is assumed that oxygen ion can not diffuse through the initial oxide layer due to its higher diffusion co-efficient - metal ions can 
only diffuse. therefore, oxide layer can only grow at the metal-gas interface, not at the metal-oxide interface. However, it is observed that in some metals 
oxygen diffuses through the initial oxide layer and reacts with metal ions. Fehler and Mott (FM) later on expanded the CM theory by introducing the 
possibility of oxygen to diffuse and contribute to the growth of oxide \cite{ch5fehlner-mott}. Rate of oxygen diffusion depends on structure of the 
initial oxide layer. If density of easy diffusion paths such as grain boundaries or dislocations are high enough in the initial oxide layer then diffusion 
of both metal as well as oxygen ions contribute to the oxide growth. In this study, HRTEM of deposited Zn--ZnO (core--shell) NCs shows that oxide scale is of polycrystalline nature and also shows the presence of dislocations at the interface. HRTEM image of the interface, having dislocations, is shown in figure \ref {ch5dislocation}(a). In order to show 
the presence of dislocations in a clear manner, the HRTEM is reconstructed by filtering the FFT pattern (figure \ref{ch5dislocation}(b)). It is seen that at
 every five or six (1 0 0) planes there is a misfit dislocation (shown by arrows). The inter-planar spacings of 0.23 nm and 0.28 nm, that are written in the 
figure, correspond to $d$-values of (1 0 0) planes of Zn and ZnO, respectively. However, the presence of dislocations resulted in observation of irregular 
Moir\'{e} fringes and fluctuation in the $d$-value due to lattice contraction and relaxation. Some experimental studies on oxidation behaviour of bulk 
Cu surfaces \cite{ch5zhou,ch5yang-kolasa} show that oxide film does not grow in a uniform layer-by-layer fashion as assumed in the CM theory, but instead, 
separate oxide islands nucleate which then grow laterally, coalesce with each other and form a thin layer over the metal surface. The process of lateral 
growth and coalescence of oxide islands results in the formation of grain boundaries. A similar mechanism might be happening during oxidation of Zn NCs 
as  oxide scale of the NCs are observed to be polycrystalline. Thus, in oxidation process of Zn NCs, diffusion of reactant ions through 
grain boundaries and dislocations might have dominated over the bulk diffusion. Hence the value of W could be much less than the one used in the 
equation (\ref{ch5cabrera}), and oxygen diffusion could have contributed significantly towards an increase in the oxide thickness. An increase in oxide thickness 
by both oxygen and metal ion diffusion is also proposed for Fe nanoparticles \cite{ch5cmwang}. 

It should be also noted that CM theory deals with oxidation behaviour of a single bulk metal surface, which is flat. But in a NC, the number of faces are more than one and corners are also involved in the oxidation process. In addition, physical and chemical properties of the material change as it reduces to the nanometer range \cite{ch5mahapatra}. In a bulk metal, thickness of the oxide is negligibly small compared with the thickness of the underlying metal. But in the case of NCs, the thickness of oxide layer, which is formed by consuming metal atoms, is comparable with the size of the original metal core.

HRTEM shown in figure \ref{ch5fig2} and \ref{ch5dislocation} are portion of  big size NCs (above $\sim$ 25 nm). The 
big size NCs (figure \ref{ch5fig1}(d)) are observed to be of geometrical shapes (namely rectangular, rhombohedral, triangular and hexagonal). The shape of a crystal, grown in laboratory, depends on two factors \cite{ch5liu}:  (i) the internal factors (bonding strength and 
symmetries of atomic arrangements) and (ii) the external factors (growth parameters and conditions). The way monomers are supplied 
around a growing crystal and the nature of foreign atoms present in the environment are two major external factors. Most of the 
crystal growth procedures used to prepare NCs do not have control over these two external factors. Therefore, original 
intrinsic shape of NCs generally get modified. In the LECBD technique, however, clusters during their growth get uniform 
supply of monomers around them as they travel along the tubular outlet. In addition, influence of foreign atoms is avoided as 
clusters are grown in a vacuum condition by using high purity materials. Thus, the shape of NCs mainly depend on internal 
factors. The symmetries displayed by the external shape of NCs are actually symmetries of its internal atomic arrangements. 
Zn belongs to the hexagonal crystal system.  Point group of hexagonal crystal system has all the symmetry elements that 
rhombohedral, orthorhombic, monoclinic, triclinic system have. Hence, deposited NCs could be of hexagonal, rhombohedral, 
orthorhombic, monoclinic, and triclinic shape. Wu {\it et al.} also observed geometrical shaped Zn NCs, produced by 
hybrid induction and laser heating method \cite{ch5wu}. In their case NCs are also produced by homogeneous nucleation process and the average size of produced NCs is more than 25 nm .

However, NCs below $\sim$25 nm are not of geometrical shapes.  It is mainly  due to effect of oxidation. Since surface 
energies and lattice constants of Zn and ZnO are different for different faces,  density of grain boundaries and dislocations  is  not same for all.  It  results in a wide variation
in  growth rate and thickness of oxide for different faces of crystal \cite{ch5cathcart,ch5lawless},  and is primarily responsible for 
distortion of the original shape. Noticeable change in shape is not observed for  big size NCs because ratio of newly formed 
oxide shell thickness to original core size is small. Observation of irregular or distorted geometrical shaped NCs is prominent for NCs of size $\sim$25 to 
15 nm (figure \ref {ch5fig1}(e)). NCs below  $\sim$15 nm, in addition to being irregularly shaped, are different in
 structure and composition. Unlike being  core-shell structure, NCs between  $\sim$15 to 5 nm are of Zn--void--ZnO type 
and smaller than that are ZnO hollow sphere type,  i.e ZnO hollow NCs. 

\begin{figure}
\includegraphics[width=8.5cm]{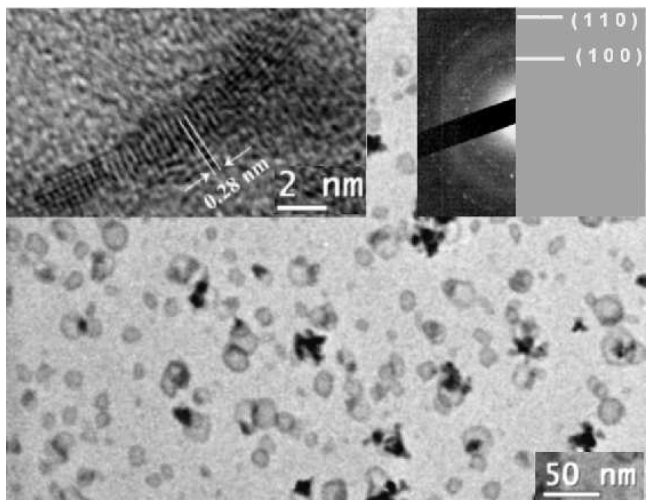}
\caption{Typical TEM micrograph of ZnO hollow NCs (sample 2). HRTEM image of shell of a hollow NC and SAED pattern obtained from hollow NCs are shown in the inset.}
\label{ch5hollow}
\end{figure}

Although diffusion of both zinc and oxygen ions contributes to the growth of oxide, their rates of diffusion are not the same.  Outward diffusion  of zinc is comparatively much higher than the inward diffusion  of oxygen, which results in accumulation of vacancies inside the metal.  When concentration of vacancies becomes higher than its equilibrium concentration, stable void forms \cite{ch5bobeth}.  It is noteworthy to mention here that vacancies diffuse much more rapidly than atoms in a
material. So distribution of vacancies inside a material is generally uniform \cite{ch5jackson}. Therefore, in NCs bigger 
than $\sim$15 nm, concentration of vacancies is not sufficient to develop supersaturation and form voids. Only in corners of some big NCs, voids are formed as a result of  local supersaturation, due to higher vacancy generation rate compared
to its diffusion during early stage of oxidation.  For NCs of size $\sim$15 to 5 nm, all Zn metal could not be converted to ZnO. Hence, a remaining amount of Zn remains in the core, with  a cap-like void and a shell of ZnO around it 
(figure \ref{ch5fig1}(f)). However,  for NCs below  $\sim$5 nm, all Zn atoms are consumed and ZnO hollow NCs are formed. The hollow ZnO NCs, Zn--void--ZnO type NCs and a big NC with a void in the corner are shown by arrow marks in figure \ref{ch5fig1}(a). It should be noted that the formation of void due to material loss is also not taken into account in the CM theory. 

The same samples are re-examined after three months of exposure to ambient. It is observed that morphology and composition of 
NCs do not change significantly. This suggests that, in agreement with the CM theory, growth of oxide stops after 
reaching certain limiting thickness. In the case of iron nanoparticles also, prepared by the gas condensation technique, the oxide shell shows a remarkable resistance to further oxidation of metal core \cite{ch5kwok}. 

\begin{figure}
\includegraphics[width=8.5cm]{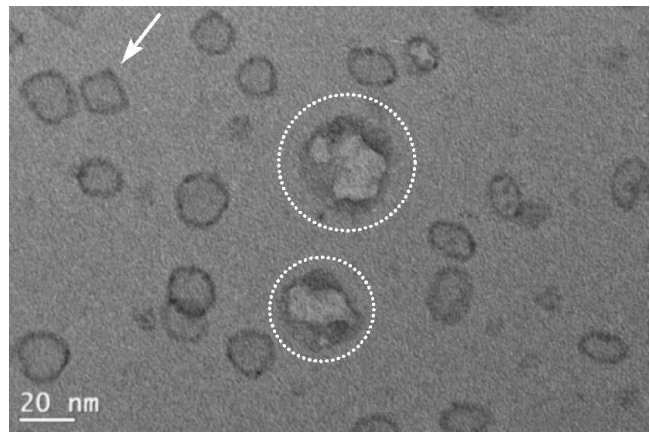}
\caption{TEM micrograph of ZnO hollow NCs at a higher magnification. Two broken NCs are marked 
by white dotted circles. An NC with distorted square-shaped void is marked by white arrow.}
\label{ch5hollow1}
\end{figure}

But when the sample 1 is further examined after three years (sample 2), it is seen that morphology and composition of NCs are significantly changed. All Zn--void--ZnO and Zn--ZnO core--shell type NCs are converted to hollow ZnO NCs. The shell thickness of hollow NCs is quite uniform (2.2 $\pm$ 0.2 nm). The $d$-spacing measured from SAED pattern matches well with those of (1 0 0) and (1 1 0) planes of ZnO. The $d$-spacing measured from HRTEM of the shell also matches well with (1 0 0) planes. The typical TEM micrograph of sample 2 at two different magnifications is shown in figure \ref{ch5hollow} and  \ref{ch5hollow1}. The SAED pattern and HRTEM micrograph are shown in the inset of figure \ref{ch5hollow}.

If a spherical Zn NC transforms completely into a ZnO  hollow sphere with its inner diameter equal to the
size of the Zn NC, then wall thickness ($t$) will be related to its inner diameter ($d_{_{Zn}}$) as follows \cite{ch5zheng}:
$$ t=\frac{d_{_{Zn}}}{2}\bigg[\bigg(1+\frac{M_{_{ZnO}}}{M_{_{Zn}}}.\frac{\rho_{_{Zn}}}{\rho_{_{ZnO}}}\bigg)^{1/3}-1\bigg]$$

Where, $M_{_{ZnO}}$ and $M_{_{Zn}}$ are molar mass of ZnO and Zn, respectively; $\rho_{_{ZnO}}$ and $\rho_{_{Zn}}$
are density of ZnO and Zn, respectively. By taking standard values of parameters \cite{ch5CRC}: $M_{_{ZnO}}=81.408$ g mol$^{-1}$,  $M_{_{Zn}}=65.38$ g mol$^{-1}$, $\rho_{_{ZnO}}=5.606$ g cm$^{-3}$, and  $\rho_{_{Zn}}=7.14$ g cm$^{-3}$ , we will obtain $ t=0.185 d_{_{Zn}} $. This means, thickness of ZnO shell will be around one fifth of the initial size of Zn NCs.  For formation of 2.2 nm thick ZnO shell, which is observed in sample 2, it is expected that Zn NCs of 11.9 nm will be completely consumed and bigger NCs will partially convert to ZnO with remaining Zn inside the core by forming Zn-void-ZnO or Zn--ZnO core--shell type NCs, as observed in sample 1. However, these type of structures are not at all observed in sample 2, though NCs bigger than 11.9 nm constituted more than fifty percent of the population in sample 1.

\begin{figure}
\begin{center}
\includegraphics*[width=8.5cm]{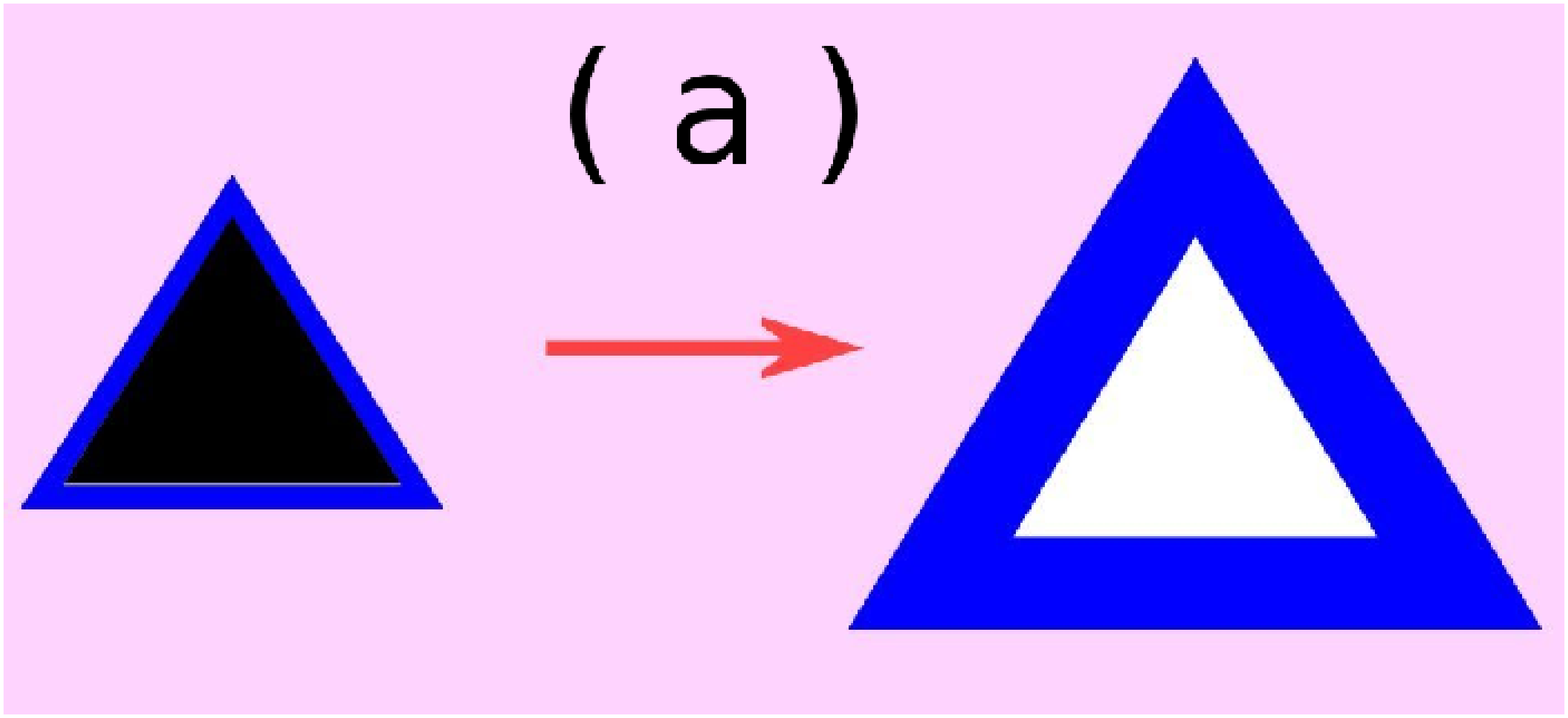}\\
\includegraphics*[width=8.5cm]{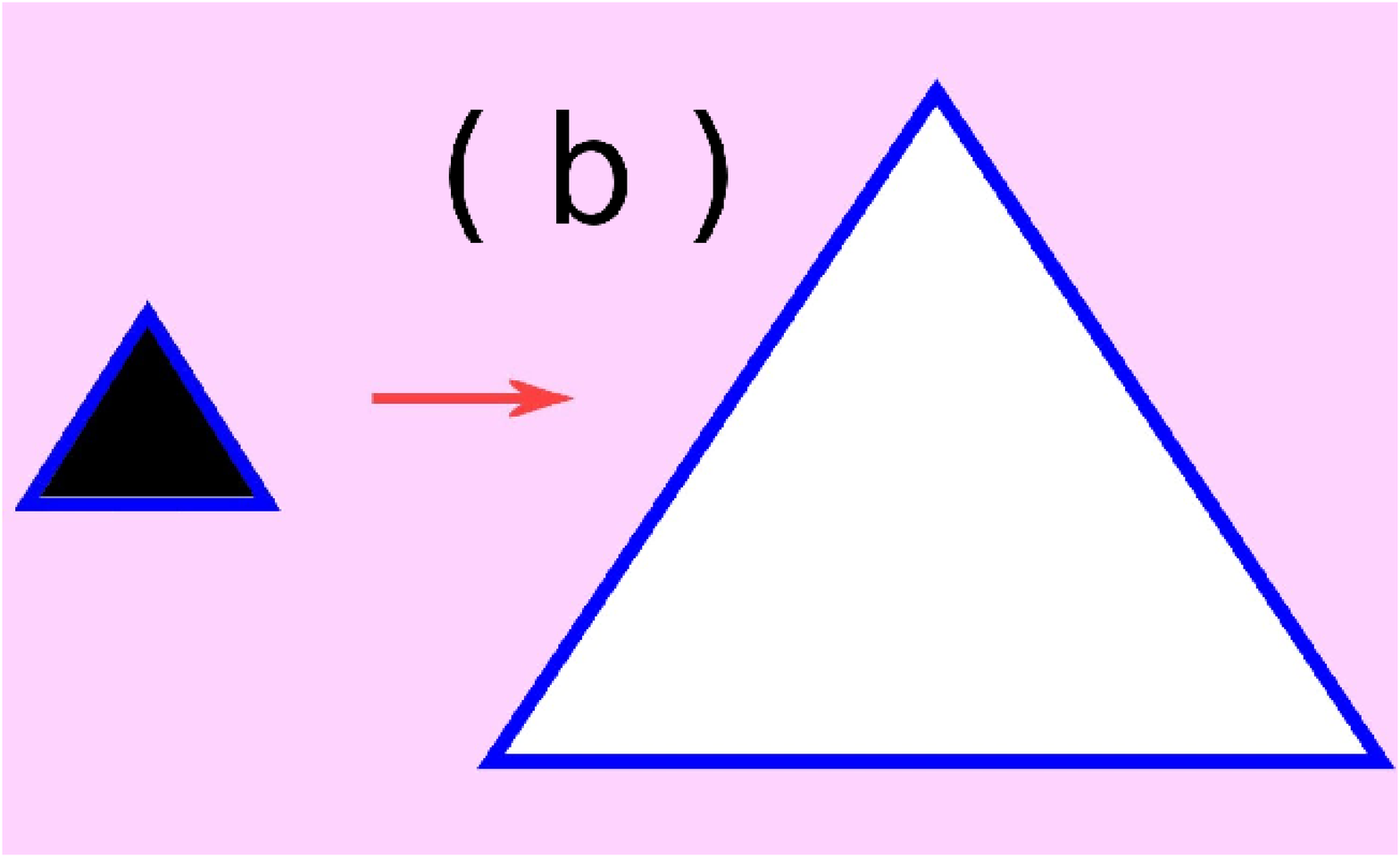}
\end{center}
\caption{{Schematic diagram of conversion mechanisms: (a) through increase in wall thickness, (b) through increase in size.}}
\label{ch5schema}
\end{figure}

\begin{figure}
\begin{center}

\includegraphics*[width=8.5 cm]{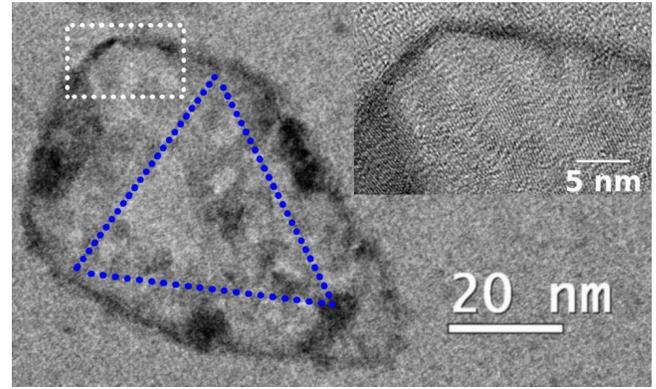}
\end{center}
\caption{TEM micrograph of a big hollow NC. Tringular shaped blue dots are marked to aid eye in getting an idea of the NC's size and probable initial shape of the Zn NC. HRTEM of the shell, marked with white dotted square, is shown in the inset.}
\label{ch5big-hollow}
\end{figure}

Zn--void--ZnO or Zn--ZnO core-shell type NCs may convert to hollow ZnO NCs by two possible mechanisms (figure \ref{ch5schema}):  
(a) through an increase in wall thickness, or (b) through an increase in size. If a spherical Zn NC, say, of 35 nm transforms to a hollow ZnO NC by 
the first mechanism, then its wall thickness will be 6.5 nm. Although NCs bigger than 35 nm were present in the sample 1, 
such thick-walled NCs are not at all observed. However, very big hollow NCs with the same 2.2 $\pm$ 0.2 nm thick wall are 
observed. Hence, it is inferred that after formation of a critical thickness, the main mechanism of conversion  is through an 
expansion in size rather than an increase in its wall thickness.  TEM micrograph of a typical big hollow NC is shown in figure \ref{ch5big-hollow}. 

\begin{figure}
\begin{center}
\includegraphics*[width=8.5 cm]{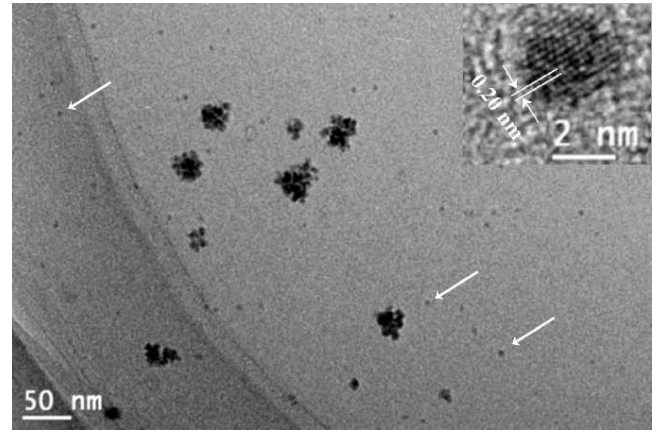}
\end{center}
\caption{A typical TEM micrograph of bursted NCs --- scattered NCs are marked by white arrows. HRTEM of a small ZnO NC is shown in the inset.} 
\label{ch5brusted}
\end{figure}

 It should be noted that size expansion is possible only if ZnO monomers are formed inside the shell. In turn, it is possible only if both zinc and oxygen ions diffuse through the oxide shell. It supplements the proposition that both oxygen and zinc diffuse through the oxide shell of Zn NCs during the process of oxidation and contribute to the growth of oxide shell. 

The following reasons might be responsible for inside material loss and creation of hollowness, even after the formation of a 
passivating oxide layer. Thermal properties of Zn and ZnO are quite different. Zn is a low melting point metal (M. P. = 693 K), whereas, ZnO is a semiconducting material and remains in solid phase up to a high temperature of
2242 K. There also exists a wide difference in the value of linear coefficient of thermal expansion, heat capacity and thermal 
conductivity of these two materials \cite{ch5porter, ch5madelung}.
In addition, thermal property changes when a material scales down to the nanometer range \cite{ch5lu}. In the present case, Zn NCs may be considered as if they are inside ZnO vacuum chambers and the phase diagram is not known in their respective value of temperature and pressure, which are also not known. Therefore, the variation in ambient temperature during different seasons of the year roughly from 285 K to 315 K may not have an insignificant 
effect on the Zn/ZnO NCs. It should be noted that 315 K  is around half of the melting point of bulk Zn. We suppose that the variation of ambient temperature creates thermal stress at metal-oxide interface and results in formation of fresh dislocations, cracks or holes in the oxide shell, through which further mass transfer occurs till all Zn converts to ZnO.

It can be noted that voids of hollow NCs are generally not spherical. Since oxide grows over the surface of metal NCs up to certain critical thickness and then expands in size through a slow and gradual process, the shape of the void is related to the initial
shape of metal NCs. The metal NCs behave like a template for the shape of the void. In sample 1 NCs above 25 nm 
were of geometrical shapes. Hence, geometrical shaped voids  are also observed in some hollow NCs in sample 2. 
One hollow NC with a rectangular shaped void is indicated by a white arrow in the figure \ref{ch5hollow1}.

Features of breakage are also seen in some NCs. Two broken NCs are marked by dotted 
white circles in figure \ref{ch5hollow1}. Gathering of big size NCs (7.4 $\pm$ 1.2 nm) at one place and scattering of 
small particles (2.2 $\pm$ 0.5 nm) around it are also observed. This gives an impression of bursting of NCs (figure \ref{ch5brusted}). Some of the scattered small ZnO NCs are marked by arrow marks. The $d$-spacing of big as well as small NCs matches well with (1 0 2) plane of ZnO. HRTEM of a small ZnO NC is shown in the inset. The contrast variation, which gives an impression of bursting of NCs, could only be seen in the case of big size NCs. It is not clear whether other small size hollow NCs are stable or broken. It should be noted that (a) compressive stress builds up during size expansion due to  internal growth of the oxide inside the shell, (b) hollow NCs, particularly with non-spherical void, are energetically unfavourable, and (c) hollow NCs with 2-3 nm thin shells may not be mechanically strong enough to sustain the atmospheric pressure. These reasons might be contributing to the instability of NCs.

\section{Conclusions}

The mechanism of oxidation proceeds in two steps: first, Zn NCs are converted to Zn--ZnO (core--shell), Zn--void--ZnO or hollow ZnO type NCs --- morphology and composition depend on initial shape and size of Zn NCs. NCs bigger than $\sim$ 15 nm become core -shell type, between $\sim$15 to 5 nm become Zn--void--ZnO type, and smaller than $\sim$5 nm become ZnO hollow sphere type. In next step, all Zn--ZnO (core--shell) type, Zn--void--ZnO type NCs convert to hollow ZnO NCs in a slow and gradual process --- the main mechanism of conversion is through expansion of size rather than increase in wall thickness. In agreement with the CM theory, thickness of oxide does not increase appreciably after reaching certain limiting value --- though the theory is not adequate enough to explain oxidation behaviour of nanometals.

\section{Acknowledgement}

The help and encouragement received from Professor S N Sahu, Professor P V Satyam, Dr S N Sarangi, Mr G Swain, Mr A K Dash, Institute of Physics, Bhubaneswar, and from Professor S R Bhattacharyya and Miss D Chowdhury, Saha Institute of Nuclear Physics, Kolkata, is gratefully acknowledged.

\end{document}